# Angled Aortic Cannulation Reduces Cerebral Embolic Transport and Adverse Wall Loading During Cardiopulmonary Bypass: A Patient-Specific Hemodynamic Study

**Nafis M. Arefin [1], Bryan C. Good [1]**

[1] Department of Biomedical Engineering
University of Tennessee, Knoxville, TN, USA

**Abstract**

Purpose: Embolic stroke during cardiopulmonary bypass (CPB) is strongly influenced by cannula-induced flow disturbances that govern emboli transport and aortic wall loading. This study quantifies how aortic cannula orientation affects embolic distribution and atherosclerotic plaque disruption risk across patient-specific, age-dependent aortic anatomies under clinical CPB conditions.

Methods: A validated computational fluid dynamics–Lagrangian particle tracking (CFD–LPT) framework was applied to four patient-specific aortic models representing pediatric, adolescent, adult, and geriatric anatomies. Two clinically relevant cannula orientations: perpendicular (90°) and angled (30°), were evaluated under varying blood viscosities (1.5–3.5 cP) and embolus sizes (0.5–2.5 mm). Aortic branch exit percentage, wall pressure, and wall shear stress (WSS) were quantified.

Results: The 30° angled cannula reduced embolic transport into the aortic branches by 18–50% compared with perpendicular cannulation, with the largest reduction observed in the geriatric model. Perpendicular cannulation produced concentrated jet impingement, resulting in significantly elevated posterior wall pressure (24%) and up to an 8-fold increase in local WSS. The pediatric anatomy exhibited the highest sensitivity, where decreasing viscosity and increasing embolus size increased branch exit by 8.4% and 25–44%, respectively, consistent with higher particle Stokes numbers and inertial decoupling from the core flow.

Conclusion: Angled cannulation redistributes jet momentum along the aortic curvature, weakens jet impingement, reduces localized pressure and wall stress, and limits embolic delivery towards the cerebral circulation. These findings provide mechanistic, patient-specific evidence supporting angled cannula orientation to reduce CPB complications.

# 1. Introduction

Cardiopulmonary bypass (CPB) is an indispensable technique in modern cardiac surgery, facilitating temporary extracorporeal circulation and oxygenation. Despite its critical role, CPB is associated with adverse complications, including the risk of embolic stroke driven by disrupted flow dynamics and emboli transport into the vasculature. Aortic cannulation during CPB, wherein oxygenated blood is reinfused in the aorta from the extracorporeal circuit, can introduce significant disturbances to physiological flow patterns [1-3]. It gives rise to turbulent structures and localized flow stagnation, creating an environment conducive to coagulation, embolus formation, and potential deviation of embolic trajectories toward the cerebral circulation [4-6].

To date, few studies have examined the detailed fluid dynamics associated with aortic cannulation [1-3, 7]. While the primary objective of CPB is to maintain physiologic perfusion that closely replicates normal cardiac function, cannula placement beneath the aortic arch can generate high-velocity jets that induce a Venturi effect, causing flow reversal in the brachiocephalic trunk. This reversal limits the delivery of oxygenated blood to brain tissue, potentially resulting in hypoxic brain injury and subsequent cognitive impairment [3]. The site, depth, and orientation of the cannula are thus critical determinants of surgical outcomes. Puthettu et al. [8], using a pseudo-cannulated aorta model, demonstrated that identifying an optimal cannulation site may significantly reduce embolic load to the brain, indicating the significance of site selection beyond surgical convenience. The ascending aorta has largely been accepted as the preferred site due to its suitability for a single incision and its capacity to accommodate larger-diameter cannulas, thereby providing higher flows based on the patient's size and metabolic needs [5, 9, 10].

Although clinical studies have attempted to evaluate the influence of cannula orientation on cerebral embolic load, their findings are often contrasting [11-14]. These discrepancies may arise from the challenges of controlling real-time surgical variables, complicating the comparability of different cases. Thus, existing evidence does not conclusively explain the influence of cannula angle on emboli trajectories, and further investigation is needed. Moreover, while some studies aim to minimize embolus initiation through cannula orientation, there is a notable absence of research examining how emboli generated in the CPB circuit are transported into the vasculature under different cannula configurations. Existing computational fluid dynamics (CFD) studies often simplify emboli as point-like or neutrally buoyant tracers, neglecting realistic mass, density, and inertial effects that influence trajectories. This includes Avrahami et al. [13], who investigated cannula designs across six orientations and 27 model cases, analyzing jet flow, shear stress, and embolic motion with point-like tracers. Similarly, Borse et al. [15] modeled emboli transport in a feline aortic arch assuming neutrally buoyant thromboemboli, despite real clots exhibiting variable densities and volumes [16, 17]. Ho et al. [7] demonstrated that a modest 15° cannula tilt toward the aortic arch reduced gaseous emboli incidence in aortic branch arteries, though embolus size variation and hemodilution effects were not considered. The present study addresses these limitations by evaluating two clinically relevant cannula orientations: perpendicular (90°) and 30° angled toward the aortic arch, with a standardized 2 cm insertion depth across varying patient anatomies. By coupling CFD with Lagrangian particle tracking (LPT) [18], this work improves emboli transport modeling by incorporating key physical parameters of mass, volume, gravitational forces, and drag laws, allowing for emboli properties to govern trajectories and for the predictions of realistic embolic pathlines rather than oversimplified streamlines.

Another critical consideration during CPB is the interaction between cannula jet flows and vulnerable atheromatous plaques in the aorta. Clinical evidence has linked these jets to plaque disruption and subsequent embolization [19-21], while CFD studies suggest that localized elevations in wall shear stress (WSS) induced by jet impingement may rupture plaques, releasing emboli into the systemic circulation [22]. Swaminathan et al. [23] reported a CPB case in which the cannula outflow was inadvertently directed toward an atheromatous plaque, resulting in embolus dislodgement. Similarly, a clinical study by Lata et al. [20] of coronary artery bypass graft patients showed that those receiving a larger CPB cannula (24Fr vs. 20 Fr) were associated with significantly fewer emboli, likely due to its lower jet velocity and impingement on plaques on the aortic wall. While clinical studies are invaluable, CFD offers a powerful means to isolate and analyze cannula-driven emboli behavior under controlled conditions [24-26]. These studies widely suggest that jet flow from the aortic cannula can instigate embolization by impacting vulnerable plaque sites along the aortic wall [24, 27, 28]. Yet, quantifying the precise mechanical criteria for plaque failure remains difficult, with most studies emphasizing that minimizing WSS helps reduce embolic risk [29-31]. However, Li et al. [35] challenged this view by showing in carotid plaque rupture models that pressure, rather than WSS, was the dominant factor initiating failure. Building on these works, the present study examines both pressure and WSS distributions under varying cannula orientations to better elucidate their combined influence on plaque disruption risk.

Another gap in the literature is the limited investigation into the impact of patient-specific geometric factors on emboli transport. Aging modifies the cardiovascular system profoundly, with arterial stiffening, wall thickening, and degeneration of elastic and collagen fibers. These changes can dilate, elongate, and curve the aorta, which in turn can alter flow patterns and pressure gradients [32, 33]. Despite this, no prior research has assessed how such anatomical variations influence embolic trajectories and stroke risk during CPB. Given that neurological morbidity rates from CPB can reach up to 30% in infants and children compared to only 5% in adults [34, 35], it is crucial to evaluate CPB parameters across all age groups, not just adults. While prior studies have examined cannula angle or emboli transport in isolation, no study has yet integrated these factors across a range of patient-specific, age-dependent anatomies. In the present study, a series of patient cases was simulated with two varying cannula angles and three blood viscosities, reflecting different levels of CPB hemodilution. Emboli of realistic size and physiological properties are then tracked to assess their transport paths and interaction with vascular geometry using LPT. By systematically examining cannula orientation, patient anatomy, and CPB conditions, this study aims to provide a comprehensive understanding of embolic injury risk during CPB. The results will offer clinically translatable recommendations for optimizing surgical practices, with the ultimate goal of reducing stroke risk during surgery.

## 2. Materials and Methods

### 2.1 Patient-Specific Aorta Model Development

Computational models of the aorta were developed for four healthy individuals representing distinct age groups: a 6-month-old pediatric, a 13-year-old adolescent, a 26-year-old adult, and a 50-year-old geriatric patient. The anatomical geometries were sourced from the Vascular Model Repository [36] and processed in *Blender* to prepare the models for simulation. As shown in Figure 1, a cylindrical CPB cannula was inserted into the ascending aorta, positioned 2 cm deep into the ascending aorta in all models except the pediatric case, where a 1 cm insertion depth was used following standard CPB protocols [37]. Cannula dimensions were selected based on patient age, body surface area, and established surgical guidelines [37-39]: 3 mm diameter for pediatric, 6 mm diameter for adolescent, and 6.67 mm diameter for both adult and geriatric patients. Two cannula orientations were implemented: one perpendicular to the ascending aorta wall (90°) and one angled 30° toward the aortic arch. Additional geometric preprocessing was performed in *Blender*, which involved removing extraneous anatomy and defining inlet and outlet boundary patches to ensure compatibility with CFD workflows. Each model included a CPB cannula as the only inlet, while the major aortic branches, the brachiocephalic artery (BCA), left common carotid artery (LCCA), left subclavian artery (LSCA), and the descending aorta (DA), were defined as outlets of the flow domain.

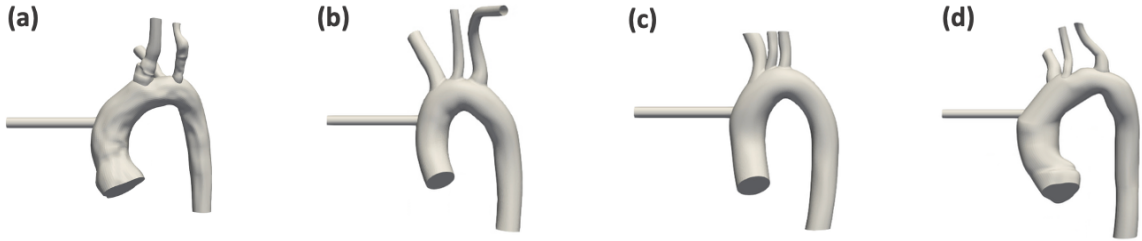

**Fig. 1** Patient-specific aortic geometries used in this study with a perpendicular (90°) CPB cannula for **(a)** pediatric, **(b)** adolescent, **(c)** adult, and **(d)** geriatric patients

### 2.2 Boundary Conditions

Boundary conditions were configured to closely replicate physiological hemodynamics observed in the aorta during CPB. Resistance-based conditions were applied to all aortic outlets to ensure accurate flow distributions, governed by Eq. 1:

$$p = p_0 + RQ \qquad (Eq.\ 1)$$

where $p$ is the outlet pressure, $p_0$ is the reference pressure, $R$ is an outlet-specific resistance, and $Q$ is the instantaneous flow rate. A reference pressure of 80 mmHg was prescribed at all outlets, consistent with the upper range of clinically targeted mean arterial pressures during CPB (MAP = 60–80 mmHg), which ensures adequate cerebral perfusion [40]. Resistance coefficients were determined iteratively for each outlet and patient model to achieve physiological flow splits of 15.7% to the BCA, 7.8% to the LCCA, 7.8% to the LSCA, and the remaining 68.7% through the DA [36]. Additionally, zero-gradient velocity conditions were imposed at all outlets to ensure consistent outflow, while no-slip conditions were enforced along all vessel walls.

### 2.3 CPB Parameters

To represent the range of clinically relevant hemodiluted blood viscosities during CPB, three different fluid viscosities were examined: 1.5, 2.5, and 3.5 cP. The CPB pump flow rate, expressed in liters per minute (LPM), was selected based on clinical standards that define flow as a function of patient body surface area (BSA), calculated using the Du Bois formula [41] (Eq. 2):

$$BSA = 0.007184 \times (Height^{0.725}) \times (Weight^{0.425}) \qquad (Eq.\ 2)$$

In accordance with these standards, distinct flow rates were assigned to each age group to reflect their respective BSA values: 0.75 LPM for the pediatric, 2.75 LPM for the adolescent, and 3 LPM for both the adult and geriatric patients. Previous studies have shown that higher CPB flow rates can substantially increase turbulence intensity and embolic dispersion. Considering that, while still adhering to physiologically relevant perfusion, flow rates within the clinical range were selected for each patient group in alignment with BSA estimates obtained from the Du Bois formula and standard cardiac index values [42, 43].

### 2.4 Emboli Parameters

To assess the impact of embolus size on transport dynamics, simulations were conducted using emboli of varying diameters. Previous studies have shown that particles smaller than 0.2 mm closely follow fluid streamlines with minimal inertial deviation [44], as such, microemboli were excluded from the present analysis. Additionally, clinical reports indicate that emboli present in the aorta typically range from 0.3 to 2.9 mm in diameter [45]. In accordance with these observations, three representative sizes: 0.5 mm, 1.5 mm, and 2.5 mm, were selected to ensure physiological relevance while capturing size-dependent transport behavior. All emboli were modeled with a constant density of 1080 kgm$^{-3}$, consistent with reported densities of emboli generated during CPB and cardiovascular device-related flows, which are slightly higher than whole blood density due to fibrin and platelet content [46, 47].

### 2.5 CFD Case Setup

Simulations were performed using *OpenFOAM* (OpenCFD, Ltd), an open-source C++ toolbox for computational continuum mechanics modeling. The Reynolds-Averaged Navier-Stokes (RANS) equations were solved under steady CPB inlet conditions using OpenFOAM's '*pisoFoam*' solver, which employs the finite volume method and the PISO algorithm to compute pressure and velocity fields [48]. A geometric–algebraic multigrid (GAMG) solver was used for pressure, while a preconditioned bi-conjugate gradient (PBiCG) solver was applied to velocity. A dynamic timestep strategy maintained a Courant number below 1, with residuals converging below 10$^{-6}$ at each time step.

To accurately resolve turbulent flow characteristics at the CPB cannula outlet, where peak Reynolds numbers reach approximately 7195, the k-ω shear stress transport (k-ω SST) turbulence model was used [49]. This model combines near-wall accuracy of the k-ω formulation with k-ε performance in the free stream, making it well-suited for transitional and separated flows [50]. The same RANS-based approach was validated in our prior work and found comparable to large eddy simulations (LES) in predicting embolic transport during CPB [51].

A mesh convergence study was conducted using three unstructured grids generated using OpenFOAM's '*snappyHexMesh*' library: coarse (~495k cells), medium (~1.04M cells), and fine (~1.49M cells). Each mesh was primarily comprised of hexahedral elements, with three additional boundary layers along the vessel walls for enhanced near-wall resolution. To assess mesh sensitivity, two representative CPB cases of the geriatric patient model were simulated: Case A with a high-viscosity fluid (3.5 cP) and 2.5 mm emboli at 3 LPM total flow rate, and Case B with a low-viscosity fluid (1.5 cP) under the same CPB conditions. Results were compared across mesh levels to evaluate numerical independence in embolic transport predictions.

### 2.6  Lagrangian Particle Tracking (LPT)

LPT is a volumetric technique that enables the tracking of discrete particles over time, even under turbulent flow regimes [41]. Embolus trajectories were simulated using *OpenFOAM's* validated LPT solver '*particleFoam*' and its '*cloudProperties*' directory. Here, each parcel was modeled as an individual embolus, with a manual injection model used to accurately control the size and initial position of each particle within the domain. To ensure statistical robustness in embolic distribution, 1000 particles were injected per simulation through the CPB cannula at the same velocity as the inflow from the CPB circuit. Particle-particle collisions were disabled, based on the assumption that an embolus would enter the circulation one at a time. Particle-wall interactions were modeled using the '*patchInteractionModel*' library, with emboli programmed to rebound upon contact with vessel surfaces. Arterial wall mechanical properties were defined using a Young's modulus of 4.66 × 10$^6$ Pa and a Poisson's ratio of 0.45 [42], applied via the '*wallModel*' utility in *OpenFOAM* to approximate compliant vascular behavior.

For all cases, flow simulations were carried out for 2.5 seconds to allow for stabilization and steady-state flow development through the aorta. Following this initialization phase, LPT simulations were continued for an additional 40 seconds to permit all injected emboli to traverse the domain and exit via one of the aortic outlets. Simulation outputs were post-processed and visualized in *ParaView* (Kitware, Inc.) (Fig. 2). During phase (i), emboli were introduced into the domain through the CPB cannula. In phase (ii), their migration toward downstream outlets was observed, and by phase (iv), the majority had exited the model through one of the prescribed outlets.

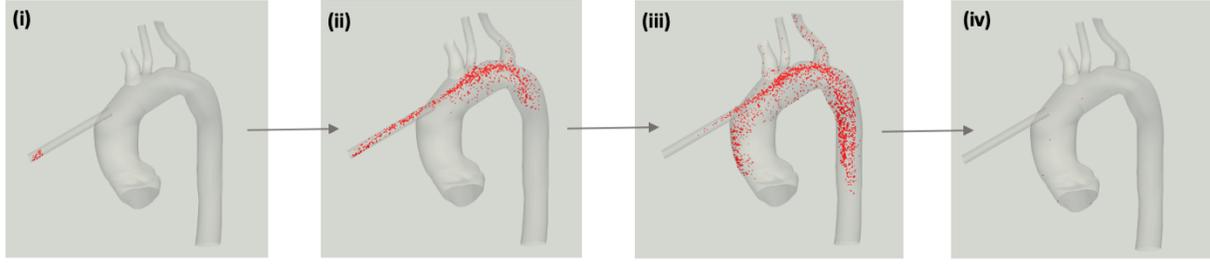

**Fig. 2** Example of computational emboli transport through the aortic arch over time in a geriatric patient

Embolus trajectories were computed using a Lagrangian framework, where their motions were governed by a system of ordinary differential equations integrated along their path. A first-order Euler integration scheme was used to update both particle position and velocity at each time step. This method enabled the resolution of translational and rotational dynamics for individual emboli throughout the simulation domain. It accounted for the dominant physical forces (Eq. 6) influencing embolic transport, including sphere drag (Eq. 7), gravitational acceleration (Eq. 9), and local pressure gradient forces (Eq. 10), allowing for physiologically accurate prediction of embolus behavior within the aorta. The translational motion of each embolus was governed by Eq. 3:

$$\frac{dx_p}{dt} = u_p \qquad (Eq.\ 3)$$

where $x_p$ is the particle's position and $u_p$ is its linear velocity. Its rotational motion was governed by Eq. 4:

$$I_p \frac{d\omega_p}{dt} = T \qquad (Eq.\ 4)$$

where $I_p$ is the particle's moment of inertia, $\omega_p$ is its angular velocity, and T is the net torque. For rigid spherical emboli, $I_p$ is calculated by Eq. 5:

$$I_p = \frac{2}{5} m_p r_p^2 \qquad (Eq.\ 5)$$

where $m_p$ is the particle's mass and $r_p$ is its radius. The net force acting on each particle was computed using Newton's second law, expressed as Eq. 6:

$$m_p \frac{du_p}{dt} = F_{drag} + F_{gravity} + F_{pressure} + \cdots \qquad (Eq.\ 6)$$

Here, the terms on the right-hand side represent the primary forces influencing embolic motion. The drag force ($F_{drag}$), which plays a dominant role in determining particle trajectories, was modeled using a mass-specific formulation appropriate for LPT solvers (Eq. 7):

$$F_{drag} = m_p \frac{\rho_p d_p^2}{18\mu} \frac{24}{C_D Re_p}(u_f - u_p) \qquad (Eq.\ 7)$$

The term $(u_f - u_p)$ in Eq. 7 represents the relative velocity between the fluid and the particle, $\rho_p$ is the particle's density, and μ is the fluid's viscosity. The drag coefficient $C_D$ is dependent on the particle Reynolds number, $Re_p$. For $Re_p$<1000, the Schiller–Naumann correlation was applied to account for both viscous and inertial effects [52] (Eq. 8a), while for $Re_p$ ≥1000, a constant drag coefficient of $C_D$ = 0.44 was adopted to approximate the drag force in the turbulent regime, based on empirical data [53] (Eq. 8b):

$$C_D = \frac{24}{Re_p}(1 + 0.15 \cdot Re_p^{0.687}),\ for\ Re < 1000 \qquad (Eq.\ 8a)$$

$$C_D = 0.44,\ for\ Re \geq 1000 \qquad (Eq.\ 8b)$$

The gravitational force ($F_{gravity}$) acting on the particle is calculated by Eq. 9, which results from spatial variations in the local fluid pressure field that is computed by Eq. 10:

$$F_{gravity} = m_p g = \frac{4}{3}\pi r_p^3 \rho_p g \qquad (Eq.\ 9)$$

$$F_{pressure} = -V_p \nabla p = -\frac{4}{3}\pi r_p^3 \nabla p \qquad (Eq.\ 10)$$

where g is gravitational acceleration and ∇p is the pressure gradient. These equations form the mathematical basis of the LPT approach in this study. By directly integrating these forces along individual embolus trajectories, the model enables physiologically accurate simulation of embolic transport through anatomically realistic CPB flow fields. This CFD–LPT framework was previously validated against in-vitro experiments using a patient-specific mock circulatory flow loop in our earlier study [51], where sample size calculation ensured adequate statistical power and the results demonstrated strong agreement with measured emboli distributions. The same validated methodology was employed in the present work.

# 3 Results

## 3.1 Grid Convergence

To evaluate the effect of mesh resolution on embolic transport predictions, embolus exit distributions through the aortic arch branches were compared across two CPB flow conditions for the geriatric patient anatomy. Case A featured a blood viscosity of 3.5 cP with 2.5 mm emboli, whereas Case B used the same viscosity with smaller 0.5 mm emboli. In Case A, the difference in embolic exits between coarse and medium meshes was 9.1%, which decreased to 4.9% between the medium and fine meshes. In Case B, the corresponding differences were 14.8% and 5.3%, respectively (Fig. 3). To quantify numerical uncertainty associated with mesh discretization, the Grid Convergence Index (GCI) was computed using the generalized Richardson extrapolation method [50]. Following Roache's recommendations for unstructured grids, a safety factor of 1.25 was applied to estimate an upper bound on the relative discretization error [54]. For Case A, the GCI values were 13.3% (medium) and 7.1% (fine), while for Case B they were 10.3% (medium) and 3.7% (fine). The progressive reduction in solution differences with mesh refinement, coupled with low GCI values, indicates that the simulations achieved asymptotic grid convergence. Therefore, the medium-resolution mesh (~1.04 million cells) was considered sufficiently refined and was adopted for all subsequent CFD simulations.

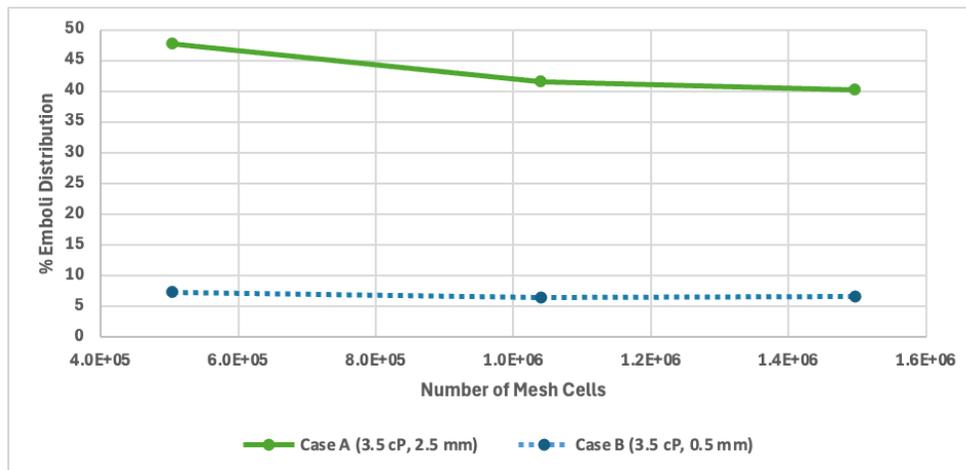

**Fig. 3** Grid convergence study of embolic transport for two cases: (A) 3.5 cP viscosity with 2.5 mm emboli, and (B) 3.5 cP viscosity with 0.5 mm emboli

## 3.2 Effect of Hemodiluted Blood Viscosity

To evaluate the impact of hemodilution on embolic transport, simulations were conducted with blood viscosities of 3.5, 2.5, and 1.5 cP (Fig. 4a). A modest but general trend of increased emboli transport to the aortic branches was observed across all patient groups as viscosity decreased. The pediatric model exhibited the most notable sensitivity to viscosity changes, with emboli transport increasing by 8.4% when viscosity was reduced from 3.5 to 1.5 cP.

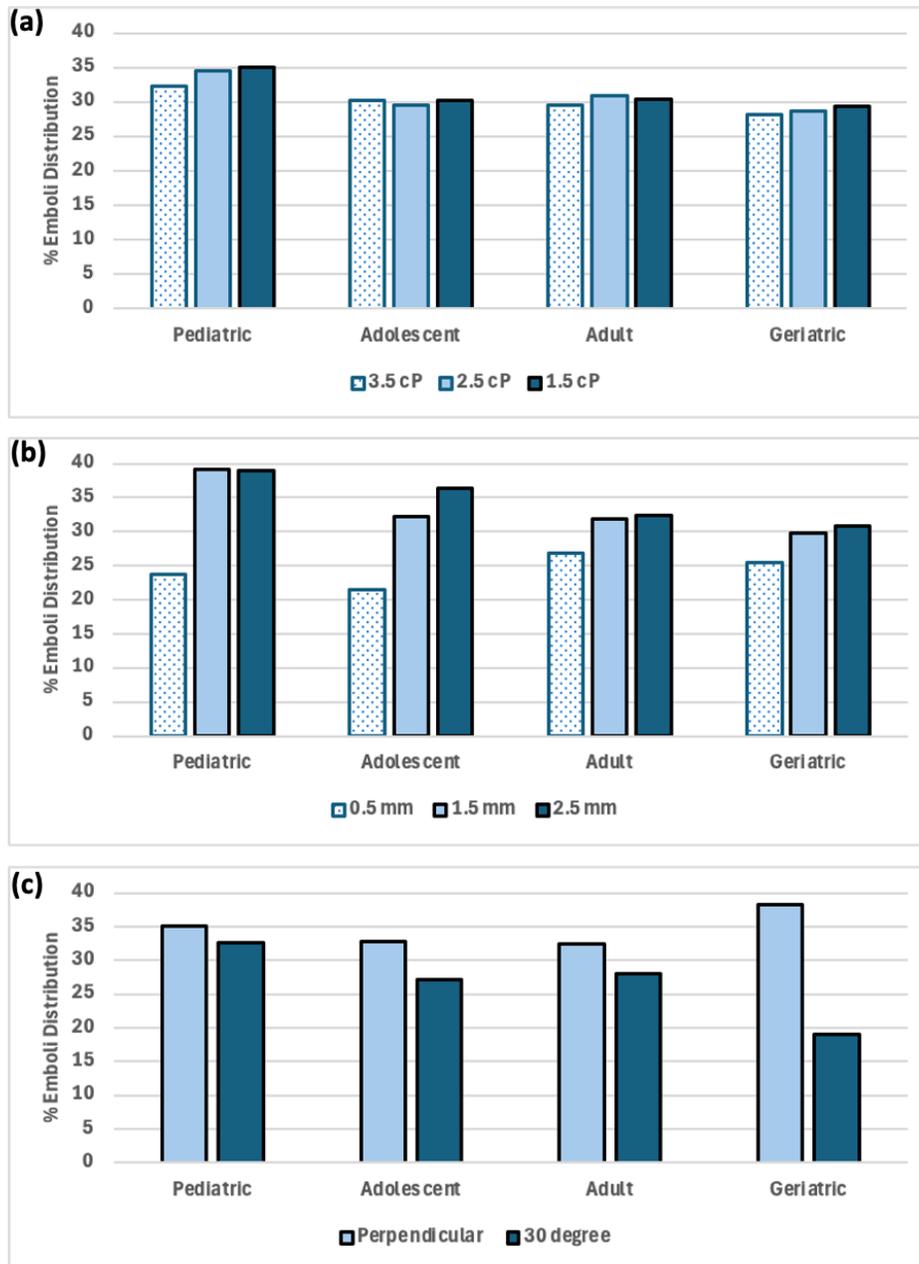

**Fig. 4** Emboli transport in the aortic branches for varying **(a)** hemodiluted blood viscosities, **(b)** emboli sizes, and **(c)** cannula orientation

### 3.3 Effect of Emboli Size

Figure 4b illustrates the influence of embolus size on transport through the aortic branches. The results show a positive correlation between embolus diameter and the likelihood of branch exit. Emboli transport increased progressively with size, with an overall rise of 25% to 44% observed when comparing 0.5 mm to 2.5 mm emboli. The most pronounced size-dependent increase occurred in the pediatric model (64.6%), while the geriatric model exhibited the smallest difference (21.3 %).

## 3.4 Effect of Cannula Orientation

The influence of cannula orientation on embolic transport was evaluated across all patient-specific aortic models by comparing a perpendicular cannula placement (90°) with a 30° angled orientation directed toward the aortic arch (Fig. 4c). In all cases, the angled cannula resulted in reduced emboli exit through the aortic arch branches. The most pronounced reduction was observed in the geriatric model, where emboli transport into the arch branches decreased by approximately 50% with the angled configuration.

Flow field differences between the two orientations are illustrated in Fig. 5 for the pediatric and geriatric models. The 90° cannula orientation produced increased flow disturbances, characterized by localized turbulence and recirculation regions, particularly within the ascending aorta, whereas the 30° angled cannula generated smoother, more streamlined flow patterns with reduced eddy formation.

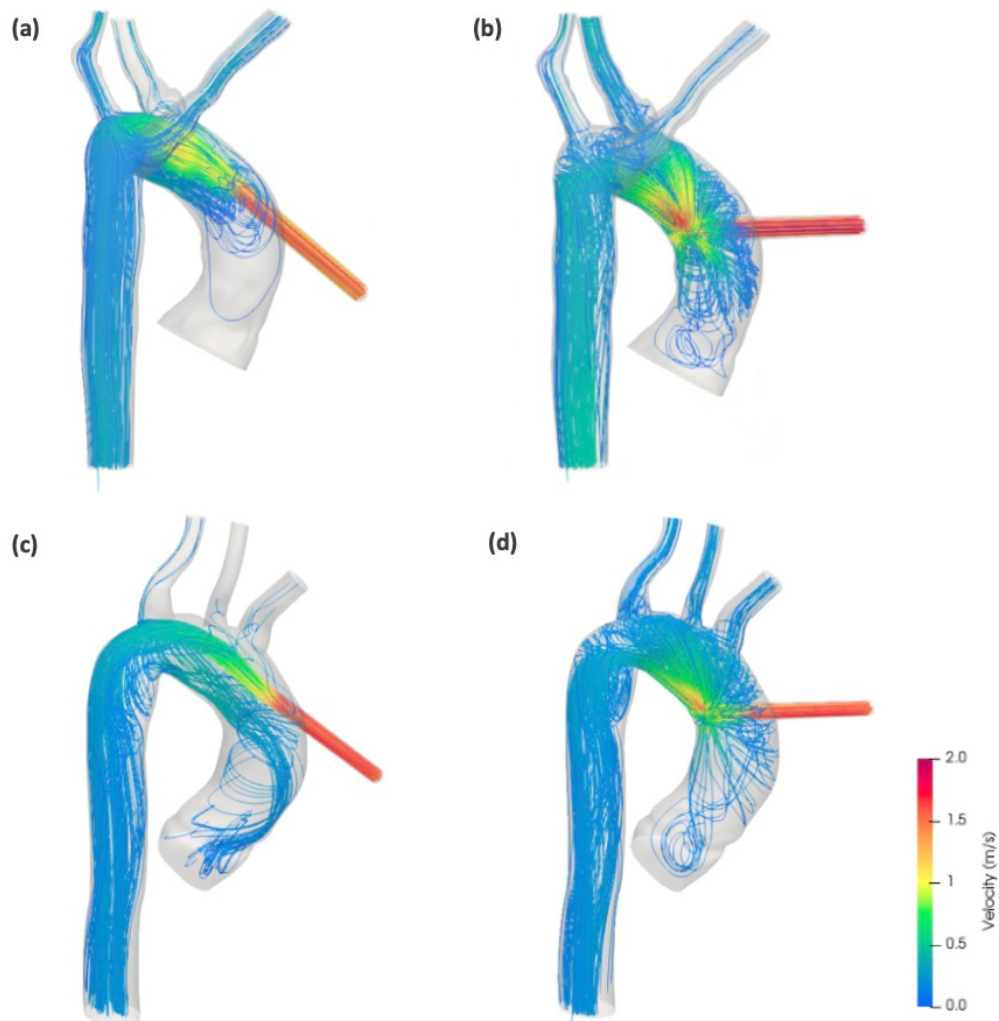

**Fig. 5** Fluid streamlines in the pediatric **(a, b)** and geriatric **(c, d)** patient aortas for cannula orientations of 30° **(a, c)** and 90° **(b, d)**

## 3.5 Atherosclerotic Plaque Disruption Risk

To evaluate the hemodynamic risk of plaque disruption, wall pressure distributions along the posterior wall of the ascending aorta were examined for both cannula orientations. The perpendicular cannula induced markedly higher localized pressures compared to the 30° angled configuration, with peak posterior wall pressure approximately 17% greater on average across all patient models. In co80ntrast, the angled cannula orientation mitigated these high-pressure regions, reducing the likelihood of adverse wall interactions. Figure 6 illustrates this behavior for the pediatric patient.

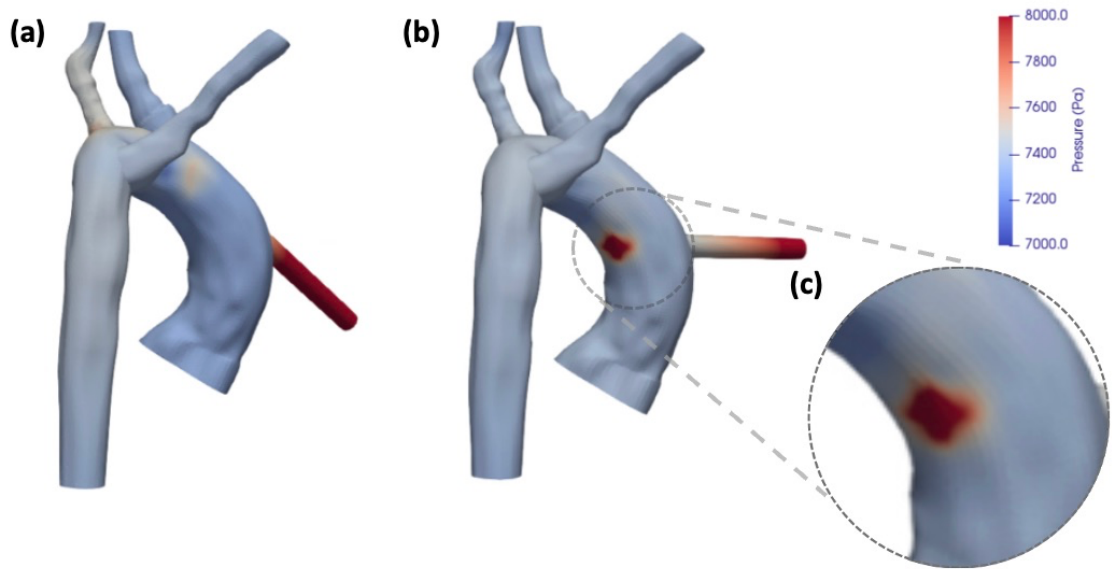

**Fig. 6** Posterior wall pressure contours for the pediatric patient: **(a)** 30° and **(b)** 90° orientations, with **(c)** magnified view highlighting region of high localized wall pressure on the posterior wall in the 90° configuration

Posterior WSS exhibited a similar dependence on cannula orientation. The perpendicular cannula generated approximately an 8-fold increase in WSS near the jet impingement region relative to the angled configuration. This effect is illustrated for the geriatric patient in Fig. 7, where the maximum WSS occurs around the point of jet impact, where the flow directly impinges on the wall (Fig. 7b). Notably, this region of elevated WSS spatially coincides with zones of increased wall pressure (Fig. 7d), indicating co-localized mechanical loading. Such combined elevations in pressure and shear are known to exacerbate plaque instability, promote disruption or fragmentation, and increase the risk of endothelial injury [55, 56]. Representative WSS and pressure contour maps illustrating these patterns in the geriatric patient case are shown in Figs. 7a–d.

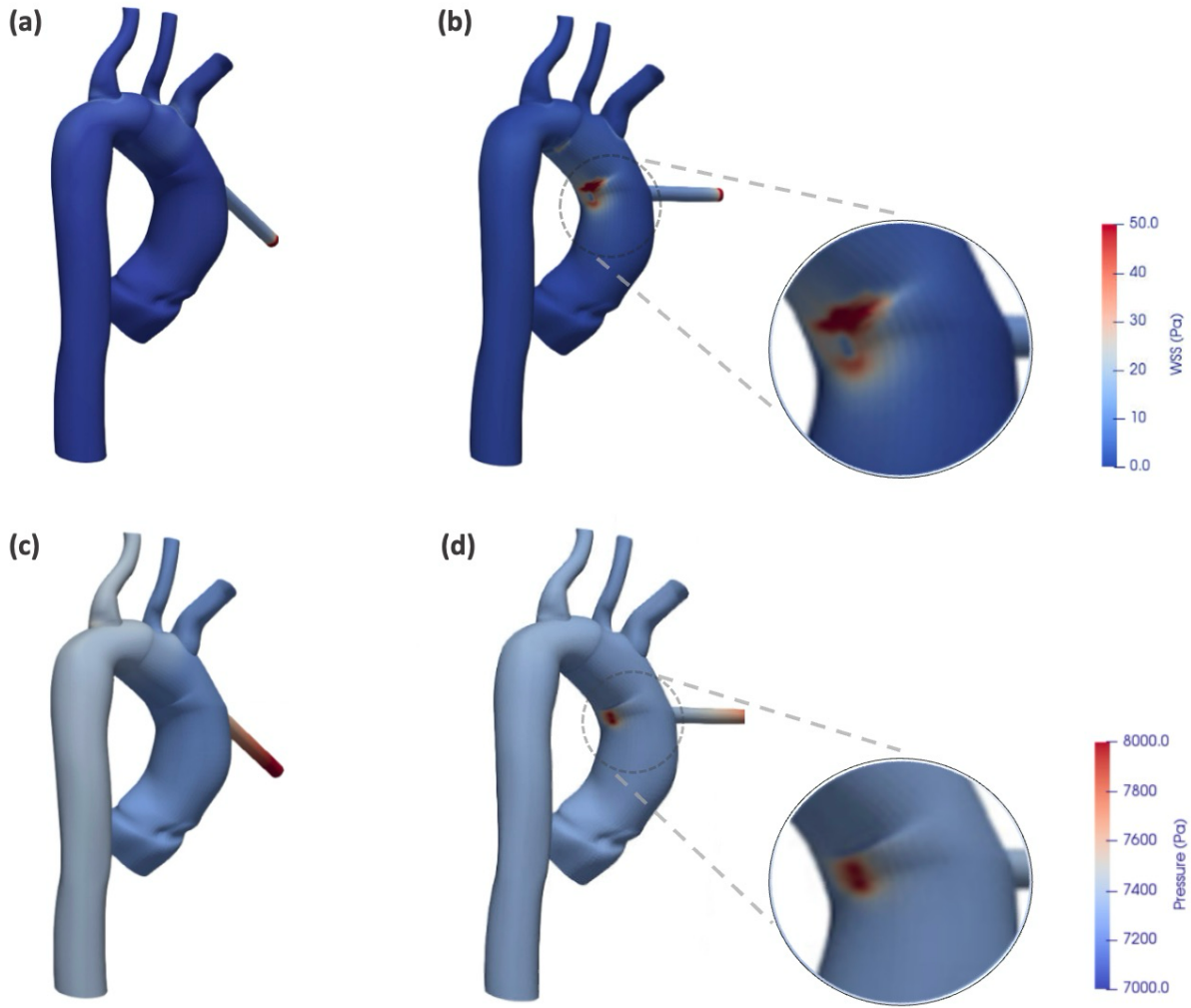

**Fig. 7** Posterior WSS **(a, b)** and pressure **(c, d)** contours in the geriatric patient aorta for cannula orientations of 30° **(a, c)** and 90°

Across all patient models and CPB conditions, Tables 1 and 2 summarize the maximum posterior wall pressures and WSSs, confirming the consistent influence of cannula orientation on local wall loading. In addition to cannula angle, blood viscosity was found to modulate both pressure and shear responses. As viscosity decreased from 3.5 cP to 1.5 cP, posterior wall pressure and WSS decreased by approximately 34.9% and 4.5%, respectively, across all models, indicating that hemodilution lessens momentum and shear transmission associated with cannula-induced jet impingement.

**Table 1:** Maximum aortic wall pressure (Pa) across age groups, blood viscosities, and cannula orientations.

| Patient | Cannula Angle (°) | Blood Viscosity (cP) | | |
|---|---|---|---|---|
| | | 3.5 | 2.5 | 1.5 |
| Pediatric | 90 | 8650 | 8350 | 8050 |
| | 30 | 7600 | 7350 | 7200 |
| Adolescent | 90 | 9300 | 9150 | 9000 |
| | 30 | 7450 | 7370 | 7280 |
| Adult | 90 | 8830 | 8650 | 8450 |
| | 30 | 7650 | 7400 | 7180 |
| Geriatric | 90 | 8250 | 8100 | 7800 |
| | 30 | 7300 | 7290 | 7270 |

**Table 2:** Maximum aortic WSS magnitude (Pa) across age groups, blood viscosities, and cannula orientations.

| Patient | Cannula Angle (°) | Blood Viscosity (cP) | | |
|---|---|---|---|---|
| | | 3.5 | 2.5 | 1.5 |
| Pediatric | 90 | 79 | 64 | 43 |
| | 30 | 20 | 18 | 17 |
| Adolescent | 90 | 92 | 78 | 54 |
| | 30 | 14 | 12 | 10 |
| Adult | 90 | 84 | 67 | 52 |
| | 30 | 13 | 11 | 10 |
| Geriatric | 90 | 126 | 98 | 63 |
| | 30 | 8 | 6 | 5 |

## 4 Discussion

This study provides a comprehensive patient-specific analysis of emboli transport and wall loading during CPB, with particular emphasis on the influence of cannula orientation, blood viscosity and embolus size. Using previously validated CFD–LPT simulations, we evaluated embolic trajectories in four anatomically distinct aortic models from patients of different age groups and quantified hemodynamic conditions associated with atherosclerotic plaque disruption risk.

**Cannula orientation:** Our results demonstrate that cannula orientation is a primary determinant of embolic distribution and wall loading. A 30° angled orientation consistently reduced emboli escape into the arch branches compared to a perpendicular orientation, with the most pronounced effect observed in the geriatric model (50% reduction), consistent with prior studies reporting orientation-dependent modulation of arch hemodynamics and embolic trajectories [3, 7]. By redirecting the jet stream obliquely, the angled configuration limited direct impingement on the posterior aortic wall, thereby both decreasing emboli entrainment into cerebral branches and mitigating localized wall pressure elevations. This effect can be explained by the alignment of momentum flux. A perpendicular jet produces a concentrated stagnation zone directly opposite the cannula tip, while an angled jet spreads its momentum along the arch, reducing focal pressure and turbulence generation. This mechanism is visually supported by the streamline patterns in Fig. 5, where the perpendicular orientation generates pronounced recirculation and flow disturbances in the ascending aorta, while the angled cannula produces smoother, more coherent streamlines aligned with the aortic curvature.

**Plaque disruption risk:** Beyond embolic transport, posterior wall pressure analysis revealed substantial differences in wall loading between cannula orientations. The perpendicular cannula generated markedly higher localized pressures (approximately 17% greater on average across all patient models) along the posterior ascending aortic wall. From a wall mechanics perspective, this configuration induces highly localized normal stresses, creating concentrated loading conditions that increase susceptibility to intimal damage and plaque destabilization [57, 58]. In contrast, the angled cannula distributes hemodynamic forces obliquely over a broader surface area, attenuating pressure peaks and resulting in weaker localized wall impingement associated with injury risk (Figs. 6 and 7).

Beyond static pressure effects, elevated posterior WSS further reinforces the potential for plaque destabilization. The substantially higher WSS generated by the perpendicular cannula indicates intensified tangential loading at the impingement zone. In the geriatric model (Fig. 7), these shear elevations coincided spatially with pressure maxima, suggesting that combined normal and tangential stresses act synergistically to weaken the local intima. Such coupled stress environments are known to accelerate endothelial erosion, promote microstructural fatigue within lipid-rich plaques, and heighten the likelihood of rupture or fragment release into circulation [55, 56].

The influence of viscosity on posterior wall pressure further highlights the interplay between CPB operating conditions and wall vulnerability. As viscosity decreased from 3.5 cP to 1.5 cP, posterior wall pressures and WSS magnitudes decreased by approximately 34.9% and 4.5%, respectively, across all models. This behavior is consistent with prior studies showing that reduced blood viscosity attenuates momentum transfer and wall shear transmission [59, 60], thereby weakening cannula jet impingement and reducing localized mechanical loading on the aortic wall, particularly downstream of cannulation sites. While hemodilution may not eliminate wall stresses entirely, this suggests that reduced viscosity can partially mitigate pressure-driven plaque disruption risk, providing an additional protective effect in susceptible patients.

**Viscosity and hemodilution:** Hemodiluted conditions amplified emboli transport into the aortic branches, with the pediatric anatomy showing the highest sensitivity. As viscosity decreased from 3.5 to 1.5 cP, emboli transport increased by up to 8.4%. This trend is consistent with prior studies demonstrating that reduced viscosity diminishes particle–fluid drag and weakens near-wall shear gradients, thereby reducing viscous damping of secondary flow structures and facilitating emboli migration into side branches [61, 62]. Under low-viscosity conditions, emboli are less constrained by viscous diffusion and more prone to deviate from central streamlines into branch flow separations, particularly in smaller, highly curved anatomies. Clinically, these findings align with reports from studies indicating that aggressive hemodilution, while often necessary to improve perfusion, may carry adverse consequences, especially in pediatric patients [51, 63].

**Embolus size:** Embolus size also exerted a strong, nonlinear influence on transport behavior. Larger emboli (2.5 mm) demonstrated up to a 44% increase in aortic branch exit compared to smaller emboli (0.5 mm), underscoring their greater tendency to deviate from central streamlines. This behavior can be explained by the Stokes number, which characterizes a particle's ability to remain coupled to the flow [61]. Larger emboli, with higher inertia, are less responsive to local fluid drag and thus more capable of crossing streamlines and entering side branches readily, a behavior consistently reported in prior embolic transport studies [51, 64]. In contrast, smaller emboli experience higher drag-to-inertia ratios, causing them to remain tightly aligned to the core flow and reducing their likelihood of branch entry. Here, the pediatric aorta exhibited the strongest size dependence, consistent with previous findings that smaller luminal diameters and sharper branch angles amplify inertial deviations of emboli [65]. In contrast, the geriatric model showed more muted size effects due to larger vessel diameters and smoother curvature. Collectively, these findings provide a mechanistic explanation for clinical observations that larger emboli are disproportionately associated with acute ischemic complications, while microemboli tend to remain entrained in central flow and traverse distally with limited deviation [44, 66].

This research has certain limitations to be acknowledged. First, emboli were modeled as rigid, spherical particles, which do not fully capture the diverse morphologies and mechanical properties observed in vivo. Real emboli may exhibit irregular shapes and viscoelastic behavior, influencing their interactions with vessel walls and altering their transport through the complex vasculature. In addition, simulations were performed under steady-state flow conditions, which, while appropriate for CPB scenarios with a clamped ascending aorta, do not fully represent the pulsatile hemodynamics present under normal physiology. Future work incorporating pulsatile inlet conditions, vessel wall compliance, and more physiologically realistic embolus models will improve the predictive strength and translational value of these simulations.

Despite these limitations, the findings offer actionable guidance for surgical practice. Angled cannula placement reduced both emboli delivery into the arch branches and pressure-driven plaque disruption risk, suggesting it as a preferable strategy in patients with known aortic atheroma. Cautious management of hemodilution may further reduce embolic dispersion, while moderating flow rates within physiologic limits can mitigate turbulence-driven emboli deviation. Together, these results highlight the need to integrate geometric, rheological, and hemodynamic considerations into CPB planning. In conclusion, this work provides the first patient-specific CFD–LPT analysis linking cannula orientation, CPB parameters, and plaque disruption risk across a range of patient-age aortic anatomies. Perpendicular cannulation amplified embolic dispersion and posterior wall pressure, while a 30° angled orientation and judicious perfusion management reduced both cerebral embolic load and plaque instability risk. These findings advance mechanistic understanding of CPB-related embolization and support optimization of cannulation strategies to enhance the safety of cardiac surgery.

**Declarations**
**Funding:** No funds, grants, or other support was received for this study.

**Competing Interests:** The authors have no relevant financial or non-financial interests to disclose.

**Authors' Contribution Statement:** N.A. performed all the computational studies, prepared the results, wrote the original draft, and reviewed the final manuscript. B.G. conceptualized and supervised the study and edited and reviewed the final manuscript.